\documentclass[12pt]{article}
\usepackage{graphicx}
\usepackage{amssymb}
\usepackage{amsmath}

\newcommand{\bear}{\begin{array}}  
\newcommand {\eear}{\end{array}}
\newcommand{\bea}{\begin{eqnarray}}   
\newcommand{\eea}{\end{eqnarray}}
\newcommand{\beq}{\begin{equation}}   
\newcommand{\eeq}{\end{equation}}
\newcommand{\bef}{\begin{figure}}  \newcommand 
{\eef}{\end{figure}}
\newcommand{\bec}{\begin{center}}  \newcommand 
{\eec}{\end{center}}
\newcommand{\non}{\nonumber}

\newcommand{\ds}{\displaystyle}

\def\lrfp#1#2#3{ \left(\frac{#1}{#2} 
\right)^{#3}}

\def\lrfp#1#2#3{ \left(\frac{#1}{#2} 
\right)^{#3}}

\begin{document}

\begin{titlepage}

\begin{flushright}
IPMU 08-0050 \\
ICRR-Report-527
\end{flushright}

\vskip 1.35cm

\begin{center}
{\large \bf
Non-Gaussianity from isocurvature perturbations
 }
\vskip 1.2cm

Masahiro Kawasaki$^{a,b}$,
Kazunori Nakayama$^a$,
Toyokazu Sekiguchi$^a$,
Teruaki Suyama$^a$ and
Fuminobu Takahashi$^b$

\vskip 0.4cm

{ \it $^a$Institute for Cosmic Ray Research,
University of Tokyo, Kashiwa 277-8582, Japan}\\
{\it $^b$Institute for the Physics and Mathematics of the Universe,
University of Tokyo, Kashiwa 277-8568, Japan}
\date{\today}

\begin{abstract}
We develop a formalism to study non-Gaussianity in both curvature and
isocurvature perturbations. It is shown that non-Gaussianity in the
isocurvature perturbation between dark matter and photons leaves
distinct signatures in the CMB temperature fluctuations, which may be
confirmed in future experiments, or possibly, even in the currently
available observational data.  As an explicit example, we consider the
QCD axion and show that it can actually induce sizable non-Gaussianity
for the inflationary scale, $H_{\rm inf} = O(10^9 - 10^{11})$\,GeV.
\end{abstract}


\end{center}
\end{titlepage}

\section{Introduction} \label{intro}
The accumulating observational data, especially the WMAP observation
of the cosmic microwave background (CMB)~\cite{Komatsu:2008hk},
provided significant support for the inflationary paradigm.  The
results of these measurements are consistent with nearly
scale-invariant, adiabatic and Gaussian primordial density
perturbations, known as the standard lore in the simple class of
inflation models.

A possible detection of the deviation from the above properties will
enable us to further constrain inflation models.  The scalar spectral
index of the power spectrum is constrained as $n_s =
0.963^{+0.14}_{-0.15}$ at 68\% C.L.~\cite{Komatsu:2008hk}, which
already excludes some inflation models.  No significant isocurvature
component has been detected so far, and the current constraint on the
ratio of the amplitudes of the isocurvature and curvature
perturbations reads, $|S/\zeta| \lesssim
0.3$~\cite{Komatsu:2008hk,Bean:2006qz}.  Recently, Yadav and Wandelt
claimed an evidence of the significant non-Gaussianity in the CMB
anisotropy data. Using the non-linearlity parameter $f_{\rm NL}$ to be
defined in the next section, their result is written as $47 < f_{\rm
  NL} < 127$ at 95\% C.L.~\cite{Yadav:2007yy}. On the other hand, the
latest WMAP five-year result is consistent with the vanishing
non-Gaussianity: $-9 < f_{\rm NL} < 111$ at 95\% C.L., including
$f_{\rm NL}=0$~\footnote{
Here we have quoted the value of $f_{\rm NL}^{\rm local}$ since we
  are interested in non-Gaussianity of the local type in this paper.
}. Interestingly, however, the likelihood distribution of the WMAP
result is biased toward positive values of $f_{\rm NL}$.  Also there
are some other studies searching for the
non-Gaussianity~\cite{Slosar:2008hx}, and it is not settled yet
whether the non-Gaussianity exists.  At the present stage, therefore,
it is fair to say that the observations are consistent with the nearly
scale invariant and pure adiabatic perturbations with Gaussian
statistics, while there is a hint of non-Gaussianity at the two sigma
level.

The standard lore on inflation is based on a simple but crude
assumption that it is only the inflaton that acquires sizable quantum
fluctuations during inflation.  Its apparent success, however, does
not necessarily mean that such a non-trivial condition is commonly met
in the landscape of the inflation theory. In fact, there are many flat
directions in a supersymmetric (SUSY) theory and the string theory. If
some of them are light during inflation, they acquire quantum
fluctuations, which may result in slight deviation from the standard
lore.

One promising candidate is provided by the theory with a Peccei-Quinn
(PQ) symmetry, which is introduced in order to solve the strong CP
problem in the quantum chromodynamics
(QCD)~\cite{Peccei:1977hh,Kim:1986ax}.  There appears a
pseudo-Nambu-Goldstone boson called axion associated with the
spontaneous breakdown of the PQ symmetry.  The axion is a light scalar
field and contributes to the cold dark matter (CDM) of the
universe~\cite{Preskill:1982cy}. In particular, the axion can have a
large isocurvature perturbation~\cite{Seckel:1985tj,Linde:1991km}.  As
for non-Gaussianity, it is known that the slow-roll inflation generally predicts a
negligible amount of non-Gaussianity, $f_{\rm NL}= O(\epsilon,
\eta)$~\cite{Acquaviva:2002ud,Maldacena:2002vr,Seery:2005wm,Yokoyama:2007uu}.  Here
$\epsilon$ and $\eta$ are the slow-roll parameters, which must be
smaller than unity for the slow-roll inflation to last long enough. In
the curvaton~\cite{Mollerach:1989hu,Linde:1996gt,Lyth:2001nq} and/or
ungaussiton~\cite{Suyama:2008nt} scenarios, however, there are light
scalars in addition to the inflaton, which can generate sizable
non-Gaussianity~\cite{Lyth:2002my,Bartolo:2003jx,Lyth:2005fi,Lyth:2006gd,
  Malik:2006pm,Ichikawa:2008iq,Beltran:2008ei, Suyama:2008nt}.
  As we will see, the axion can also induce sizable non-Gaussianity.

In this paper we point out that, if an isocurvature component
possesses some amount of non-Gaussianity, it is transferred to the
non-Gaussianity of the curvature perturbation, resulting in a possibly
large value of $f_{\rm NL}$.  If there are no other light scalar
fields than the inflaton, the resultant density perturbations are
necessarily adiabatic and almost Gaussian. As mentioned before, this
may not be the case in the presence of many flat directions.  Suppose
that there is a light scalar that acquires quantum fluctuations during
inflation. Then its fluctuations produce isocurvature component.  If
the scalar decays into radiation, the isocurvature perturbation is
converted into the adiabatic one. This is exactly what occurs in the
curvaton and/or ungaussiton scenarios.  However, as far as
non-Gaussianity is concerned, the light scalar having large
fluctuations needs not decay. 
Even if such a light field is not responsible for the total curvature perturbation, 
it can still provide a source of large non-Gaussianity.
This interesting possibility was noted in 
Refs.~\cite{Linde:1996gt,Bartolo:2001cw,Boubekeur:2005fj}. 
In this paper we have systematically studied the non-Gaussainity from the
isocurvature perturbations and how it exhibits itself in the CMB anisotropy.
Interestingly enough, we have found that the resultant non-Gaussianity induced by the isocurvature perturbations
has distinctive signatures in the CMB, which should be distinguished
from that in the curvaton and ungaussiton scenarios.

This paper is organized as follows.  In Sec.~\ref{formalism}, a
general formalism to study non-Gaussianity including isocurvature
perturbations is presented.  In Sec.~\ref{sec:temp}, we compute the
bispectrum of the temperature fluctuations arising from
non-Gaussianity in the isocurvature perturbations.  In
Sec.~\ref{axion} the formalism is applied to the case of the axion and
it is shown that the axion can induce large non-Gaussianity while
leaving a certain amount of the CDM isocurvature perturbation.
Sec.~\ref{conclusion} is devoted to discussion and conclusions.

\section{Non-linear isocurvature perturbation} \label{formalism}

\subsection{Definition of the isocurvature perturbation}

Let us consider cosmological perturbations of multicomponent fluids
labeled by $i = 1, \dots, n$. We assume that the density perturbations
originate from fluctuations of scalar fields generated during
inflation.

We write the spacetime metric as
\begin{equation}
	ds^2=-{\mathcal N}^2 dt^2 +a^2(t)e^{2\psi} \gamma_{ij} \left ( dx^i + \beta^i dt \right )
	\left ( dx^j + \beta^j dt \right ),
\end{equation}
where ${\mathcal N}$ is the lapse function, $\beta_i$ the shift
vector, $\gamma_{ij}$ the spatial metric, $a(t)$ the background scale
factor, and $\psi$ the curvature perturbation.  On sufficiently large
spatial scales, the curvature perturbation $\psi$ on an arbitrary slicing at
$t=t_f$ is expressed by~\cite{Lyth:2004gb}
\begin{eqnarray}
	\psi (t_f,{\vec x})=N(t_f,t_i:{\vec x})-\log \frac{a(t_f)}{a(t_i)},
	 \label{curvature1}
\end{eqnarray}
where the initial slicing at $t=t_i$ is chosen in such a way that the
curvature perturbations vanish (flat slicing).  Here $N(t_f,t_i:{\vec
  x})$ is the local $e$-folding number, given by the integral of the local expansion along the worldline
${\vec x}={\rm const.}$ from $t=t_i$ to $t=t_f$.

We denote by $\zeta$  the curvature perturbation $\psi$ evaluated on the slice where the
total energy density is spatially uniform (uniform-density slicing). In a similar fashion,
we also introduce $\zeta_i$ to denote the curvature perturbation on the slice
where $\rho_i$ is uniform ($\delta \rho_i=0$ slicing).  Then, from
Eq.~(\ref{curvature1}), $\zeta_i$ is related to $\zeta$ by the gauge
transformation
\begin{eqnarray}
	\zeta_i=\zeta + \Delta N_i, \label{curvature2}
\end{eqnarray}
where $\Delta N_i$ is the $e$-folding number measured from the
uniform-density slicing to the $\delta \rho_i=0$ slicing, both
slicings corresponding to the same background time.   
If each component of the fluids does not exchange its energy with the
others, $\zeta_i$ are known to remain constant
for the scales larger than the horizon~\cite{Lyth:2004gb}.

Let us define $\delta \rho_i$ as the perturbation of $\rho_i$ on the
uniform-density slicing. That is, 
\beq
\rho_i(N,{\vec x})  \;=\; \rho_{i0} + \delta \rho_i(N,{\vec x}),
\eeq
where $\rho_{i0}$ is the energy density of the $i$-th fluid in the background spacetime,
and $N = N({\vec x})$ defines the uniform density slicing. 
Then $\Delta N_i$ is related to $\delta
\rho_i$ by the following equation,
\begin{eqnarray}
	\rho_i (N+\Delta N_i,{\vec x})=\rho_i (N,{\vec x})-\delta \rho_i(N,{\vec x}), \label{gauge1}
\end{eqnarray}
where the l.h.s and r.h.s are evaluated on the $\delta \rho_i =0$
slicing and on the uniform-density slicing, respectively.  Assuming 
$\delta \rho_i/\rho_{i0} \ll 1$ and $\Delta N_i \ll 1$, we can solve this 
equation with respect to $\Delta N_i$ up to the second order in
$\delta \rho_i$:
\begin{eqnarray}
	\Delta N_i \; \simeq\;-\frac{\delta \rho_i}{\rho_{i0}'} -
	\frac{\rho_{i0}''}{2\rho_{i0}'}{\left( \frac{\delta \rho_i}{\rho_{i0}'} \right)}^2
	+\frac{\delta \rho_i \delta \rho_i'}{{\rho_{i0}'}^2}, \label{gauge2}
\end{eqnarray}
where
the prime denotes the derivative with respect to $N$.\footnote{
	If the $i$-th fluid has vanishing homogenous value, i.e., if it is produced predominantly by
	the quantum fluctuations, $\delta \rho_i/\rho_{i0}$ as well as $\Delta N_i$ is no longer small.
	In the example of axion which we discuss later, 
	this problem can be avoided by considering the density contrast of the {\it total} CDM sector.
}
Hence $\zeta_i$ can be written as
\begin{eqnarray}
	\zeta_i \;\simeq\; \zeta-\frac{\delta \rho_i}{\rho_{i0}'}-
	\frac{\rho_{i0}''}{2\rho_{i0}'}{\left( \frac{\delta \rho_i}{\rho_{i0}'} \right)}^2
	+\frac{\delta \rho_i \delta \rho_i'}{{\rho_{i0}'}^2}. \label{gauge3}
\end{eqnarray}

We define the (non-linear) isocurvature perturbation between the
$i$-th fluid and the $j$-th one as \cite{Wands:2000dp}
\begin{eqnarray}
	S_{ij} \equiv 3(\zeta_i-\zeta_j). \label{iso1}
\end{eqnarray}
Using Eq.~(\ref{gauge3}),
$S_{ij}$ can be written as
\begin{eqnarray}
	S_{ij}\;\simeq\;3\left[ -\frac{\delta \rho_i}{\rho_{i0}'}+\frac{\delta \rho_j}{\rho_{j0}'} 
	-\frac{\rho_{i0}''}{2\rho_{i0}'}{\left( \frac{\delta \rho_i}{\rho_{i0}'} \right)}^2
	+\frac{\rho_{j0}''}{2\rho_{j0}'}{\left( \frac{\delta \rho_j}{\rho_{j0}'} \right)}^2 
	+\frac{\delta \rho_i \delta \rho_i'}{{\rho_{i0}'}^2}
	-\frac{\delta \rho_j \delta \rho_j'}{{\rho_{j0}'}^2}
	\right]. \label{iso2}
\end{eqnarray}
If we neglect the second order terms, $S_{ij}$ reduces to the
well-known form. If the $i$-th fluid fluctuates in the same way as the
$j$-th one, i.e., $\Delta N_i = \Delta N_j=0$, the isocurvature
perturbation between the two, $S_{ij}$, vanishes.  All the
isocurvature perturbations vanish if there is only the adiabatic
perturbation, that is, if all $\Delta N_i$ vanish.

We assume that the density perturbations originate from the
fluctuations of light scalar fields during inflation. Then $\delta
\rho_i$ can be expanded as~\footnote{
	   Note that $\{ \delta \rho_i\}$ are subject to a constraint $\sum_i\delta \rho_i=0$
	   on the uniform density slicing since.
}
\begin{eqnarray}
	\delta \rho_i\;=\;\rho_{i,a} \delta \phi^a+\frac{1}{2} \rho_{i,ab} \delta \phi^a \delta \phi^b+\cdots, \label{iso3}
\end{eqnarray}
where $\delta \phi_a$ is the quantum fluctuation of a light scalar
$\phi_a$ on the initial flat slicing at $t=t_i$.  
We choose the
initial time $t_i$ slightly after the cosmological scales of interest exit the
Hubble horizon, since the above formulation is valid for the
superhorizon modes.  We assume that the scalar fields, $\{\phi_a\}$,
have quadratic potential and
behave like free fields and do not have any sizable interactions
during inflation.  In particular, their masses are assumed to be
lighter than the Hubble parameter during inflation.  
In this case, the higher order terms in Eq.~(\ref{iso3}) are safely neglected.
Then, to a good approximation, $\delta \phi_a$ is given by the Gaussian variable.
In general, the coefficients that appear in the right-hand side of Eq.~(\ref{iso3}) depend on the slicing on which they are evaluated.
When evaluating the coefficients, we need to choose an appropriate uniform-density slicing.
	For example, if $\rho_i$ denotes the energy density of the axion, those coefficients are 
	  easily evaluated on the uniform density slicing
	when the axion starts to oscillate.	
	If $\rho_i$ is the energy density of a particle produced by the decay of a scalar field,
	those coefficients include the information from the onset of the filed oscillation
	to its decay. Thus case-by-case calculations are required. 
Substituting Eq.~(\ref{iso3}) into Eq.~(\ref{iso2}),
$S_{ij}$ can be written in the form
\begin{eqnarray}
	S_{ij}\;\simeq\;S_{ij,a} \delta \phi^a+\frac{1}{2} S_{ij,ab} \delta \phi^a \delta \phi^b
	\label{Sij}
\end{eqnarray}
with
\begin{eqnarray}
S_{ij,a}&\equiv&-3 \left( \frac{\rho_{i,a}}{\rho_{i0}'}- \frac{\rho_{j,a}}{\rho_{j0}'} \right),\\
S_{ij,ab}&\equiv&-3\left( \frac{\rho_{i,ab}}{\rho_{i0}'} -\frac{\rho_{j,ab}}{\rho_{j0}'} \right)
-3\left( \frac{\rho_{i0}'' \rho_{i,a}\rho_{i,b}}{{\rho'}_{i0}^3} -\frac{\rho_{j0}'' \rho_{j,a}\rho_{j,b}}{{\rho'}_{j0}^3}\right)\non\\
&&+6 \left(\frac{\rho_{i,a} \rho_{i,b}'}{\rho_{i0}'^2}-\frac{\rho_{j,a} \rho_{j,b}'}{\rho_{j0}'^2}\right).
\end{eqnarray}

For simplicity, we assume that the masses of $\{\phi_a\}$ are
negligible, and the fluctuations are independent to each other. Then
the correlation functions are given by the following form,
\begin{equation}
	\langle \delta \phi^a_{\vec k_1} \delta \phi^b_{\vec k_2} \rangle\;=\;(2\pi)^3\,P_{\delta \phi}(k_1)
	\delta(\vec k_1+\vec k_2)\delta^{ab}
	\label{phiphi}
\end{equation}
with
\beq
P_{\delta \phi}(k) \;\simeq\; \frac{H_{\rm inf}^2}{2k^3},
\eeq
where $k$ denotes the comoving wavenumber, and 
$H_{\rm inf}$ is the Hubble parameter during inflation.  For
later use, we also define the following:
\beq
\Delta_{\delta \phi}^2 \;\equiv\; \frac{k^3}{2\pi^2} P_{\delta \phi}(k) \simeq \lrfp{H_{\rm inf}}{2\pi}{2}.
\eeq

\subsection{Bispectrum of the isocurvature perturbations}
We define the power spectrum and bispectrum of $S_{ij}$ as
\begin{eqnarray}
	\langle S_{ij {\vec k_1}} S_{ij {\vec k_2}} \rangle \;\equiv\;
	{(2\pi)}^3 P_{Sij} (k_1) \delta ({\vec k_1}+{\vec k_2}),  \label{Psij}
\end{eqnarray}
and
\begin{eqnarray}
	\langle S_{ij {\vec k_1}} S_{ij {\vec k_2}} S_{ij {\vec k_3}} \rangle \;\equiv\;
	{(2\pi)}^3 B_{Sij} (k_1,k_2,k_3) \delta 	({\vec k_1}+{\vec k_2}+{\vec k_3}).
\end{eqnarray}
Here and in what follows no summation is taken over the indices $i$ and $j$,
while we sum over the repeated indices $a, b, c, \dots$.
Using (\ref{Sij}) and (\ref{phiphi}), the power spectrum can be expressed as
\begin{equation}
	P_{Sij}(k)\;=\;S_{ij,a}S_{ij,a}P_{\delta \phi}(k) 
	+\frac{1}{2}S_{ij,ab}S_{ij,ab} \int \frac{d^3 {\vec k'}}{(2\pi)^3}P_{\delta \phi}(k')
	P_{\delta \phi}(|{\vec k}-{\vec k'}|),
\end{equation}
where we have regarded $\delta \phi^a$ as a Gaussian variable.
After performing the integration, we obtain
\begin{equation}
	P_{Sij}(k)\;=\;\left[ 
		S_{ij,a}S_{ij,a}+S_{ij,ab}S_{ij,ab} \Delta_{\delta \phi}^2 \ln(kL)
	\right] P_{\delta \phi}(k),
\end{equation}
where we have introduced an infrared cutoff $L$ that is taken to be of order of the present 
Hubble horizon scale~\cite{Lyth:1991ub,Boubekeur:2005fj,Lyth:2007jh}. 
Similarly, the bispectrum can be written as
\bea
B_{Sij}(k_1,k_2,k_3) &= &S_{ij,a}S_{ij,b}S_{ij,ab}
		\left[ P_{\delta \phi}(k_1)P_{\delta \phi}(k_2)
		+({\rm 2~perms}) \right]\non\\
	&& +  \,S_{ij,ab}S_{ij,bc}S_{ij,ca}    
		\int \frac{d^3 \vec k'}{(2\pi)^3}P_{\delta \phi}(k') P_{\delta \phi}(|\vec k_1-\vec k'|)
		P_{\delta \phi}(|\vec k_2+\vec k'|).
\eea
In the squeezed configuration in which one of the three wavenumbers is much smaller than the other two (e.g. $k_1 \ll k_2,k_3$), 
it is approximately given by
\bea
	B_{Sij}(k_1,k_2,k_3) &\simeq &\left[ 
		S_{ij,a}S_{ij,b}S_{ij,ab}+S_{ij,ab}S_{ij,bc}S_{ij,ca}\Delta_{\delta \phi}^2\ln (k_bL)
	\right] \non\\
	&&\times\left[ P_{\delta \phi}(k_1) P_{\delta \phi}(k_2)+
	P_{\delta \phi}(k_2) P_{\delta \phi}(k_3)+P_{\delta \phi}(k_3) P_{\delta \phi}(k_1)\right], \label{b}
\eea
where $k_b\equiv {\rm min}\{ k_1,k_2,k_3\}$.

Let us define the non-liearity parameter of the isocurvature perturbations, $f_{Sij}$, as
\begin{eqnarray}
	B_{Sij} (k_1,k_2,k_3) \;\equiv\;f_{Sij} [ P_{Sij}(k_1) P_{Sij}(k_2)+P_{Sij}(k_2) P_{Sij}(k_3)+P_{Sij}(k_3) P_{Sij}	(k_1)]. \label{fSij}
\end{eqnarray}
We can see that $f_{Sij}$ is not very sensitive to the wavenumbers.
If $S_{ij}$ is dominated by the linear terms in $\delta \phi_a$ (see (\ref{Sij})),
$f_{Sij}$ becomes independent of the wavenumbers, and given by
\begin{eqnarray}
	f_{Sij}\;\simeq\;\frac{S_{ij,b}S_{ij,c}S_{ij,bc}}{ {(S_{ij,a}S_{ij,a})}^2},  \label{fSij-1}
\end{eqnarray}
 {\it for generic configurations of the wavenumbers}.
Even if the quadratic part dominates, i.e.,  $S_{ij}\simeq 1/2\,S_{ij,ab}\delta \phi^a \delta \phi^b$,
its dependence is only logarithmic in the squeezed configuration:
\begin{equation}
	f_{Sij}\;\simeq\; \frac{1}{\Delta_{\delta \phi}^2 \ln(k_bL)}
		\frac{S_{ij,ab}S_{ij,bc}S_{ij,ca}}{ {(S_{ij,ab}S_{ij,ab})}^2},  \label{fSij-2}
\end{equation}
where we have approximated as $\ln(k_iL)\simeq \ln(k_bL)$ for $i=1,2,3$.
For a generic configuration, the dependence may become more involved.
Nevertheless, we expect that such dependence is also mild for the scales of interest,
 based on the dimensional arguments.

\subsection{``$f_{\rm NL}$" from $f_{Sij}$}

In many literatures, the non-linearity parameter $f_{\rm NL}$ is used to measure
the non-Gaussianity of the adiabatic perturbations (for example,
see Ref.~\cite{Bartolo:2004if} and references therein). 
We adopt the following conventional definition of $f_{\rm NL}$,
\begin{equation}
	\langle \zeta_{ {\vec k_1}} \zeta_{ {\vec k_2}} \zeta_{ {\vec k_3}} \rangle \;\equiv\;
	\frac{6}{5}f_{\rm NL}(2\pi)^3\delta(\vec k_1+\vec k_2+\vec k_3)
	\left[ P_\zeta(k_1)P_\zeta(k_2)+2~{\rm perms.} \right].  
	 \label{defnl}
\end{equation}
Now we would like to relate $f_{\rm NL}$ to $f_{Sij}$ defined by Eq.~(\ref{fSij}).
This is a non-trivial task since we are considering the non-Gaussianity of the isocurvature 
perturbations.  
To be definite, we hereafter consider the CDM isocurvature perturbations.
It can be extended to the other types of the isocurvature perturbations 
in a similar way.

We write the adiabatic perturbations originating from the inflaton fluctuations
as $\zeta^{\rm inf}$. Using the formula (\ref{curvature1}), it can be expressed
as
\beq
\zeta^{\rm inf} \;=\; N_\phi\, \delta \phi,
\eeq
where $\delta \phi$ denotes the fluctuation of the inflaton $\phi$, and
$N_\phi$ is the derivative of the local $e$-folding number with respect
to $\phi$. We assume that primordial inflation does not generate large non-Gaussianity.
In fact,  it was shown that the non-linearlity parameter generated during slow-roll inflation 
is at most of the order of the slow-roll parameters, and hence below the sensitivity
of the Planck satellite~\cite{Maldacena:2002vr,Seery:2005wm,Yokoyama:2007uu}.
Then the three-point function of  $\zeta^{\rm inf}$ is approximately given by
\beq
\langle \zeta^{\rm inf}_{ {\vec k_1}} \zeta^{\rm inf}_{ {\vec k_2}} \zeta^{\rm inf}_{ {\vec k_3}} 
	\rangle \;\simeq\; 0.
\eeq

Let us now define the CDM isocurvature perturbation in the universe after the reheating as 
\beq
S \;\equiv\; 3(\zeta_{\rm CDM}-\zeta_r),
\eeq
where $\zeta_{{\rm CDM}(r)}$ is the curvature perturbation on the slicing where
the energy density of the CDM(radiation) is spatially uniform.
We assume that the CDM is always decoupled from the radiation, 
so that $\zeta_{\rm CDM}$  as well as $\zeta_r$ are time-independent.  
Note that $\zeta$ is not necessarily conserved in the presence of the
isocurvature perturbation. 

When the universe is dominated by the radiation,
the curvature perturbation on the super-horizon scales is given by
$\zeta =\zeta_r$. We assume that the curvature perturbation at that time
is originated solely from the inflaton, i.e.,  $\zeta_r = \zeta^{\rm inf}$.
In the matter dominated era, we have $\zeta = \zeta_{\rm CDM}$.
Hence $\zeta$ in the matter dominated era can be written as
\begin{equation}
	\zeta\;= \;\zeta^{\rm inf}+\frac{1}{3}S, ~~~~{\rm (matter~dominated~era)},
\label{zeta-S}
\end{equation}
where we have assumed that the curvature perturbation mainly comes from
the inflaton and the other fields contribute only to the isocurvature perturbations.
Note that this relation holds to any orders in the perturbative expansion.

We assume that  the isocurvature perturbation is uncorrelated with the 
primordial curvature perturbation, i.e.,
\beq
\langle \zeta^{\rm inf}_{\vec k_1} S_{\vec k_2} \rangle \;=\;0.
\eeq
The three-point function of the curvature perturbation is then evaluated as
\begin{equation}
	\langle \zeta_{ {\vec k_1}} \zeta_{ {\vec k_2}} \zeta_{ {\vec k_3}} \rangle = 
	\frac{1}{27}\langle S_{\vec k_1}S_{\vec k_2}
	S_{\vec k_3} \rangle. \label{3point}
\end{equation}
From (\ref{fSij}), (\ref{defnl}) and (\ref{3point}), $f_{\rm NL}$ is related to $f_S$ as follows,
\begin{equation}
\begin{split}
	\frac{6}{5}f_{\rm NL} &=
		\frac{1}{27}f_S
		\frac{P_S(k_1)P_S(k_2)+({\rm 2~perms})}
		{P_\zeta(k_1)P_\zeta(k_2)+({\rm 2~perms})} \\
	&= \frac{1}{27 N_\phi^4}f_S\left[ 
		S_{,a}S_{,a}+S_{,ab}S_{,ab} 
		\Delta_{\delta \phi}^2 \ln(kL)
	\right]^2.
	 \label{fNL}
\end{split}
\end{equation}
It should be noted that the above relation between $f_{\rm NL}$ and $f_S$ is valid
only for the large scales which enter the horizon after the matter-radiation equality.
This however helps us to get a feeling of the non-Gaussianity produced from
the isocurvature perturbation.

\section{CMB Temperature Fluctuations}  \label{sec:temp}
In this section, we calculate how the non-Gaussianity of the isocurvature perturbation
exhibits itself in the CMB anisotropy, following the notations used in 
Ref.~\cite{Bartolo:2004if,Komatsu:2001rj}. In particular, it contributes to
the bispectrum of the CMB temperature fluctuations, which may be
observable in the future observations.

We introduce spherical harmonic coefficients of the temperature
anisotropy arising from the CDM isocurvature perturbations as
\begin{eqnarray}
a_{\ell m}^{\rm (iso)}\;=\;\int d{\vec n} ~\frac{\Delta T^{\rm (iso)}({\vec n})}{T}\, Y_{\ell m}^* ({\vec n}). 
\label{tem1}
\end{eqnarray}
In order to relate the primordial fluctuations to the CMB temperature anisotropy,
one needs to multiply the transfer function. Precisely speaking, one has to use
a non-linear version of the transfer function, which also induces a certain amount of
non-Gaussianity. Based on the dimensional grounds, however,
such secondary non-Gaussianity is expected to be much smaller than the value 
currently hinted by the observation. Since we are interested in the relatively large 
 primordial non-Gaussian features in the CMB anisotropy, we can 
neglect the intrinsic non-linear property in the transfer function.
We therefore use the linear transfer function,
$g^{\rm (iso)}_{T \ell}(k)$, defined by
\begin{equation}
\Theta^{\rm (iso)}_\ell({\vec k}) \;\equiv\; g^{\rm (iso)}_{T \ell}(k)S_{\vec k},\label{eq:deftrans}
\end{equation}
where $\Theta^{\rm (iso)}_{\ell}({\vec k})$ is the multipole moment of CMB temperature anisotropy:
\begin{equation}
\frac{\Delta T^{\rm (iso)}({\vec n})}{T}\;=\;\int \frac{d^3k}{(2\pi)^3}
\sum_\ell i^\ell (2\ell+1)\Theta^{\rm (iso)}_\ell({\vec k})
P_\ell({\hat {\vec k}}\cdot{\vec n}). \label{eq:phbf}
\end{equation}
Here $P_\ell$'s are the Legendre polynomials.
Using Eqs.~(\ref{eq:deftrans}) and (\ref{eq:phbf}), 
$a_{\ell m}^{\rm (iso)}$ can be written as
\begin{eqnarray}
a_{\ell m}^{\rm (iso)}\;=\;4\pi i^\ell \int \frac{d^3 k}{ {(2\pi)}^3 }\, g_{T \ell}^{\rm (iso)} (k) \,Y^*_{\ell m}({\hat {\vec k}}) \,S_{\vec k}. \label{tem2}
\end{eqnarray}

The anglar power spectrum of $a_{\ell m}^{\rm (iso)}$ is defined by
\beq
\langle a_{\ell m}^{\rm (iso)} a_{\ell' m'}^{\rm (iso)*} \rangle\; \equiv\;C_\ell^{\rm (iso)}\delta_{\ell \ell'} \delta_{mm'}.
\eeq
Using (\ref{tem2}), we obtain
\begin{eqnarray}
C_\ell^{\rm (iso)}\;=\;\frac{2}{\pi} \int_0^\infty dk~k^2 {( g_{T \ell}^{\rm (iso)}(k) )}^2 P_S (k). \label{tem3}
\end{eqnarray} 
Here and in what follows, we use $k=0$ to show the lower limit of the integration interval,
although it is set to be the infrared cutoff $L^{-1}$ in the actual calculation.
The angular bispectrum of $a_{\ell m}^{\rm (iso)}$ is defined by
\begin{eqnarray}
\langle a_{\ell_1 m_1}^{\rm (iso)}a_{\ell_2 m_2}^{\rm (iso)}a_{\ell_3 m_3}^{\rm (iso)}\rangle 
\;\equiv\; B^{{\rm (iso)} m_1 m_2 m_3}_{~~~~~\ell_1 \ell_2 \ell_3}. \label{tem4}
\end{eqnarray}
Statistical isotropy divides the angular bispectrum into
the following form,
\begin{eqnarray}
B^{{\rm (iso)} m_1 m_2 m_3}_{~~~~~\ell_1 \ell_2 \ell_3}\;=\;{\cal G}^{m_1 m_2 m_3}_{\ell_1 \ell_2 \ell_3} b_{\ell_1 \ell_2 \ell_3}^{\rm (iso)}. \label{tem5}
\end{eqnarray}
Here ${\cal G}^{m_1 m_2 m_3}_{\ell_1 \ell_2 \ell_3}\equiv \int d{\vec n}~Y_{\ell_1 m_1}({\vec n})Y_{\ell_2 m_2}({\vec n})Y_{\ell_3 m_3}({\vec n})$ (Gaunt integral) and $b_{\ell_1 \ell_2 \ell_3}^{\rm (iso)}$ is the
reduced bispectrum, on which we will focus in the following.

Substitutiing (\ref{tem2}) into (\ref{tem4}), we obtain 
\begin{eqnarray}
&&b_{\ell_1 \ell_2 \ell_3}^{\rm (iso)}= \frac{8}{\pi^3}\int_0^\infty dr~r^2 \int_0^\infty dk_1~k_1^2 \int_0^\infty dk_2~k_2^2 \int_0^\infty dk_3~k_3^2 \nonumber \\
&&\hspace{15mm}\times g_{T \ell_1}^{\rm (iso)}(k_1)j_{\ell_1}(k_1r)g_{T \ell_2}^{\rm (iso)} (k_2) j_{\ell_2}(k_2r)g_{T \ell_3}^{\rm (iso)} (k_3)j_{\ell_3}(k_3r) B_S (k_1,k_2,k_3) \nonumber \\
&&\hspace{11mm}=\frac{8}{\pi^3}\int_0^\infty dr~r^2 \int_0^\infty dk_1~k_1^2g_{T \ell_1}^{\rm (iso)}(k_1)j_{\ell_1}(k_1r) P_S(k_1)\int_0^\infty dk_2~k_2^2g_{T \ell_2}^{\rm (iso)}(k_2)j_{\ell_2}(k_2r) P_S(k_2) \nonumber \\
&&\hspace{15mm}\times \int_0^\infty dk_3~k_3^2 g_{T \ell_3}^{\rm (iso)} (k_3)j_{\ell_3}(k_3r)~f_S (k_1,k_2,k_3) +(2~{\rm perms}), \label{tem6}
\end{eqnarray}
where $j_\ell (x)$ is the spherical Bessel function, and
we have used Eq.~(\ref{fSij}) in the last equality.
From the discussion below Eq.~(\ref{b}),
we have seen that $f_S$ depends on the three wavenumbers at most logarithmically,
i.e. the dependence is rather weak.
Therefore, we approximate $f_S$ as the constant and write the bispectrum as
\begin{eqnarray}
b_{\ell_1 \ell_2 \ell_3}^{\rm (iso)} \;\simeq\;
 f_S \int_0^\infty dr~r^2 \left( b_{{\rm L} \ell_1}^{\rm (iso)}(r) b_{{\rm L} \ell_2}^{\rm (iso)}(r) b_{{\rm NL} \ell_3}^{\rm (iso)} (r) +2 ~{\rm perms.} \right), \label{tem7}
\end{eqnarray}
where $b_{{\rm L} \ell}^{\rm (iso)}$ and $b_{{\rm NL} \ell}^{\rm (iso)}$ are
defined by
\begin{eqnarray}
&&b_{{\rm L} \ell}^{\rm (iso)}(r) \;\equiv\; \frac{2}{\pi} \int_0^\infty dk~k^2g_{T \ell}^{\rm (iso)}(k)j_{\ell}(kr) P_S(k), \label{eq:bl_lin}\\
&&b_{{\rm NL} \ell}^{\rm (iso)}(r) \;\equiv\; \frac{2}{\pi} \int_0^\infty dk~k^2 g_{T \ell}^{\rm (iso)} (k)j_{\ell}(kr).
\label{eq:bl_nonlin}
\end{eqnarray}

In a similar way, we can define $C_\ell$ and $b_{\ell_1 \ell_2 \ell_3}$ as the angular power spectrum
and the reduced bispectrum of the {\it total} temperature anisotropy including both the adiabatic and the isocurvature
contributions. We  define the non-linearity parameter $f_{\rm NL}^{\Delta T}$ as
\begin{eqnarray}
	b_{\ell_1 \ell_2 \ell_3} \;\equiv\; \frac{6}{5}f_{\rm NL}^{\Delta T} 
	\int_0^\infty dr~r^2 \left( b_{{\rm L} \ell_1}(r) b_{{\rm L} \ell_2}(r) b_{{\rm NL} \ell_3}(r) +2 ~{\rm perms.} \right)
	\label{bladi}
\end{eqnarray}
with
\begin{eqnarray}
	&&b_{{\rm L} \ell}(r) \equiv \frac{2}{\pi} \int_0^\infty dk~
	k^2g^{\rm (adi)}_{T \ell}(k)j_{\ell}(kr) P_\zeta(k), \\
	&&b_{{\rm NL} \ell}(r) \equiv \frac{2}{\pi} \int_0^\infty dk~k^2 g^{\rm (adi)}_{T \ell}(k)j_{\ell}(kr),
\end{eqnarray}
where $g^{\rm (adi)}_{T \ell}(k)$ is the transfer function for the adiabatic perturbations.
Note that it is $f_{\rm NL}^{\Delta T}$ that is directly related to the CMB observations.\footnote{
	A rigorous procedure to constrain non-Gaussianity from isocurvature perturbations 
	using observational data can be done in a way demonstrated in Ref.~\cite{Komatsu:2003iq}.
	Here we simply use $f_{\rm NL}^{\Delta T}$ as a representative value 
	which characterizes the non-Gaussianity in the CMB anisotropy. 
}
If only the adiabatic perturbation exists, $f_{\rm NL}^{\Delta T}$
coincides with $f_{\rm NL}$ defined by Eq.~(\ref{defnl}).
However, the relation between $f_{\rm NL}^{\Delta T}$ and $f_{\rm NL}$ gets rather involved
when the isocurvature perturbation mainly contributes to the bispectrum 
while the power spectrum is dominated by the primordial adiabatic contribution, i.e., 
$C_\ell \simeq C_\ell^{\rm (adi)}$ and $b_{\ell_1 \ell_2 \ell_3} \simeq b_{\ell_1 \ell_2 \ell_3}^{\rm (iso)}$.
In particular, it should be noted that $f_{\rm NL}^{\Delta T}$ sensitively depends on $(\ell_1, \ell_2, \ell_3)$.

Table~\ref{tablefNL} summarizes the non-linearity parameters which we have defined so far,
$f_S, f_{\rm NL}$ and $f_{\rm NL}^{\Delta T}$.
Given a model, $f_S$ is easily calculated by Eqs.~(\ref{fSij-1}) and (\ref{fSij-2}).
Once we know $f_S$ we can obtain $f_{\rm NL}$ through the relation (\ref{fNL}).
But the most relevant quantity directly related to the CMB observations {\it is} $f_{\rm NL}^{\Delta T}$
and we evaluate it numerically in the following.

\begin{table}
\begin{tabular}{ l | l| c }
	Non-linearity parameter & Related to & Definition \\  \hline
	$f_S$ & 3-point function of isocurvature perturbation & Eq.(\ref{fSij}) \\ 
	$f_{\rm NL}$ & 3-point function of curvature perturbation & Eq.(\ref{defnl}) \\
	$f_{\rm NL}^{\Delta T}(\ell_1, \ell_2, \ell_3)$ & 3-point function of temperature perturbation & Eq.(\ref{bladi}) \\
\end{tabular}
\caption{Non-linearity parameters.}
\label{tablefNL}
\end{table}

\subsection{Sachs-Wolfe approximation}
For the low multipoles, typically smaller than $\ell \sim 10$,
the temperature anisotropy comes mainly from the Sachs-Wolfe effect,
\begin{equation}
\left(\frac{\Delta T}{T}\right)_{\rm SW} \;=\; -\frac{1}{5}\zeta^{\rm inf} - \frac{2}{5}S. \label{tem8}
\end{equation}
From this equation,
the transfer function in the Sachs-Wolfe regime can be written as
\begin{eqnarray}
g_{T \ell}^{\rm (iso)}(k)\;=\;-\frac{2}{5}j_\ell(kr_*), \label{eq:SWtrans}
\end{eqnarray}
where $r_*$ is the comoving distance to the last scattering surface from us.
Then the reduced bispectrum becomes
\begin{eqnarray}
b_{\ell_1 \ell_2 \ell_3}^{\rm (iso)} \;\simeq\; -\frac{5}{2}f_S \left( C_{\ell_1}^{\rm (iso)} C_{\ell_2}^{\rm (iso)}+2~{\rm perms.} \right). \label{tem9}
\end{eqnarray}
As we will see in the next section,  the above expression (\ref{tem9})
is not that precise since $b_{{\rm NL}\ell}(r)$ has non-negligible contributions from  
smaller scales beyond the Sachs-Wolfe plateau. It is still useful however
to understand the large amplitude of the bispectrum
from isocurvature  perturbations with non-Gaussianity. 
Here we continue with this approximation and derive a relation between $f_{\rm NL}^{\Delta T}$ and $f_{\rm NL}$
which is valid up to a $O(1)$ numerical factor,
and leave detailed discussions for the next subsection.

Under this approximation, $f_{\rm NL}^{\Delta T}$ defined in Eq.~(\ref{bladi}) is expressed as
\begin{eqnarray}
b_{\ell_1 \ell_2 \ell_3} = -6f_{\rm NL}^{\Delta T} \left( C_{\ell_1} C_{\ell_2}+2~{\rm perms.} \right).
\end{eqnarray}
We require that the adiabatic perturbations dominate the power spectrum, i.e.,  $C_\ell \simeq C_\ell^{\rm (adi)}$.
When only the adiabatic perturbations exist,
$f_{\rm NL}^{\Delta T}$ exactly coincides with  $f_{\rm NL}$~\cite{Bartolo:2004if}.
However, in the presence of the isocurvature perturbations,
$f_{\rm NL}^{\Delta T}$ is different from $f_{\rm NL}$.
If the bispectrum is dominated by the isocurvature perturbation, i.e., $b_{\ell_1 \ell_2 \ell_3} \simeq b_{\ell_1 \ell_2 \ell_3}^{\rm (iso)}$,
$f_{\rm NL}^{\Delta T}$ can be written as
\begin{eqnarray}
f_{\rm NL}^{\Delta T}\simeq\frac{5}{12} \frac{C_{\ell_1}^{\rm (iso)} C_{\ell_2}^{\rm (iso)}+2~{\rm perms.}}{C_{\ell_1}^{\rm (adi)} C_{\ell_2}^{\rm (adi)}+2~{\rm perms.}} f_S. \label{tem10}
\end{eqnarray}
Combining this relation with Eq.~(\ref{fNL}) yields
\begin{equation}
	f_{\rm NL}^{\Delta T}\simeq216 f_{\rm NL}. \label{eq:f_TNL}
\end{equation}
This is a quite important result. When the non-Gaussianity comes from the isocurvature
perturbations, the non-Gaussian features appearing in the CMB anisotropy is greatly 
enhanced in the Sachs-Wolfe plateau. This is because the isocurvature perturbations
make more contributions to
the CMB power spectrum at low multipoles than the adiabatic perturbations.
As we will see below, full treatment of the transfer functions beyond the Sachs-Wolfe plateau 
will change the numerical factor $216$ in Eq.~(\ref{eq:f_TNL}) to about $100$ .

\subsection{Acoustic scales}
In this section we present more detailed discussion on the 
bispectrum from the non-Gaussian isocurvature perturbations,
especially focusing on their differences from 
the adiabatic perturbations.
To study the features in the bispectrum at small angular scales
beyond the Sachs-Wolfe plateau, 
we have numerically calculated $b_{{\rm L} \ell}(r)$, $b_{{\rm NL} \ell}(r)$ and the reduced bispectrum $b_{\ell_1 \ell_2 \ell_3}$
 using transfer functions $g_{T \ell}(k)$ from the {\tt CAMB} code~\cite{Lewis:1999bs}.
Throughout this section we adopt the flat SCDM model
and assume a  set of cosmological parameters
$(\Omega_b=0.05,\, \Omega_c=0.95, \,h = 0.5)$, where
$\Omega_{b(c)}$ is the density parameter of the baryon(CDM),
and $h$  is the Hubble parameter in units of $100$km/sec/Mpc.
For simplicity, we neglect the tilt of the power spectra, 
$P_S(k)$ and $P_\zeta(k)$  in Eq.~(\ref{fNL}). Then both $P_S(k)$ and $P_\zeta(k)$
are proportional to $k^{-3}$.


\begin{figure}[t]
 \begin{center}
   \includegraphics[width=1.0\linewidth]{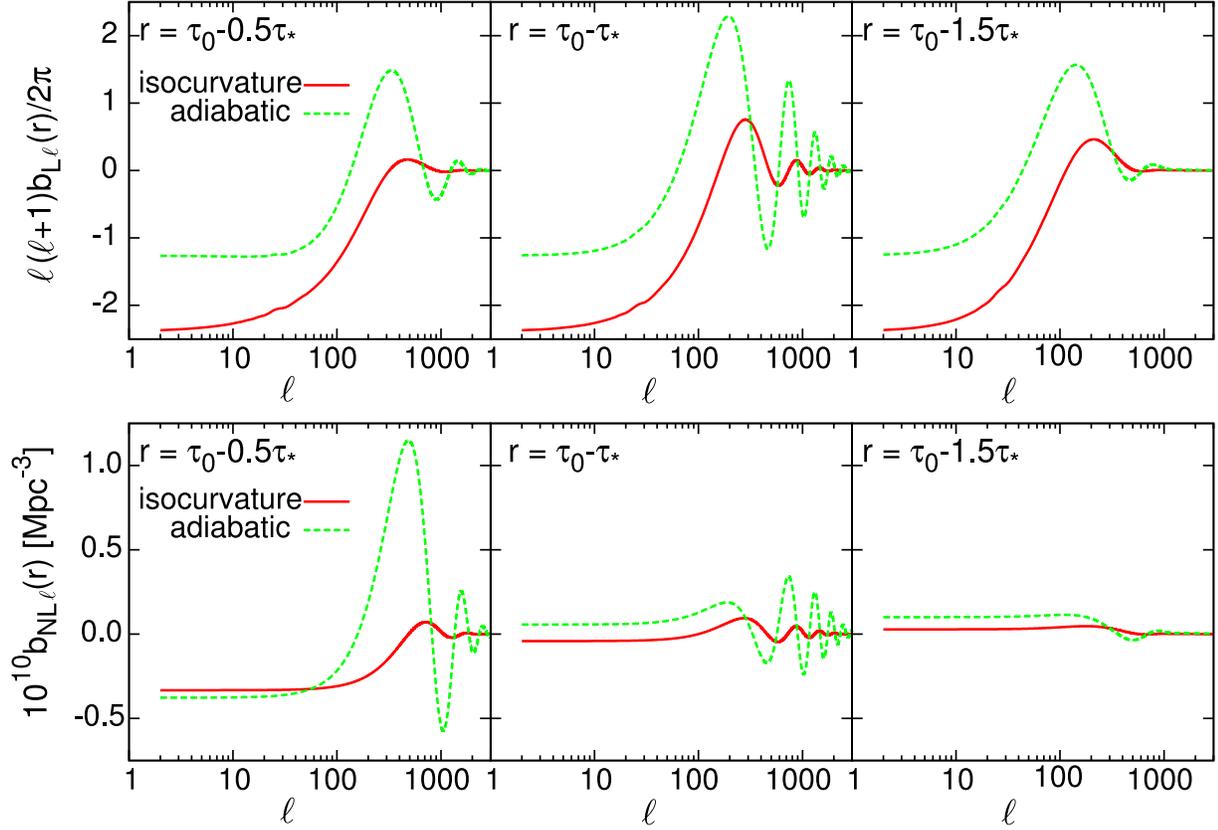}
   \caption{$b_{{\rm L} \ell}(r)$ (top) and $b_{{\rm NL} \ell}(r)$ (bottom) from numerical calculation. 
   We show the cases of isocurvature initial conditions (solid red line) 
   and adiabatic conditions (dashed green line).   Here we have assumed $P_S(k)=P_\zeta(k)$.
   From left to right panels, $r$ is set to be $r = \tau_0-0.5\tau_*,\ \tau_0-\tau_*$ and $\tau_0-1.5\tau_*$, 
   respectively. Note that $b_{{\rm L} \ell}$ is a dimensionless quantity, while $b_{{\rm NL} \ell}$ has 
   the dimensionality of $(\mbox{length})^{-3}$. 
   }
   \label{fig:bl}
 \end{center}
\end{figure}


In Fig.~\ref{fig:bl} we show the numerical results of
$b_{{\rm L} \ell}(r)$ and $b_{{\rm NL} \ell}(r)$.
If we take $r$ largely different  from $r_*=\tau_0-\tau_*$, which is the comoving distance
from us at $\tau = \tau_0$ to the last scattering surface at $\tau = \tau_*$,
both $b_{{\rm L} \ell}$ and $b_{{\rm NL} \ell}$ get suppressed.
This is because $g_{T \ell}(k)\sim j_\ell(kr_*)$ and $j_\ell(kr)$
in the integrant of Eqs.~(\ref{eq:bl_lin}-\ref{eq:bl_nonlin}) would oscillate with different 
frequencies, making contributions in a destructive way.
Therefore the signature of the primordial non-Gaussianity mostly comes from $r\simeq r_*$.
In Fig.~\ref{fig:bl} we have taken several values of $r$ around $r_*$.

Let us first consider $b_{{\rm L} \ell}(r)$ shown in the upper panels of Fig.~\ref{fig:bl}. 
We  notice that $b^{\rm (iso)}_{{\rm L} \ell}$ is 
roughly twice as large as $b^{\rm (adi)}_{{\rm L} \ell}$ in the amplitude at large angular 
scales ($\ell \lesssim 10$) for any values of $r\sim r_*$.
This behavior is similar to the angular power spectra $C_\ell$. 
It  can be easily understood by noting that the Sachs-Wolfe effect leads to
\begin{equation}
g^{\rm (iso)}_{T \ell}(k)\simeq 2g^{\rm (adi)}_{T \ell}(k). \label{eq:ratiot}
\end{equation}
At smaller angular scales $b_{{\rm L} \ell} (r)$'s represent the acoustic oscillations. 
Note that the phase of oscillations are different by $\pi/2$ between the
isocurvature and adiabatic perturbations. 
This is another similarity of $b_{{\rm L} \ell}(r)$ with $C_\ell$.

On the other hand, the situation with $b_{{\rm NL} \ell}$  is slightly different. 
Athough $b_{{\rm NL} \ell}$'s also give similar flat spectra at the large angular scales, 
they have some differences from $b_{{\rm L} \ell}(r)$.
One of them is that the 
ratio of the amplitudes of $b_{{\rm NL} \ell}$ for isocurvature and adiabatic initial conditions
differs with $r$. This is because $b_{{\rm NL} \ell}(r)$ receives more contribution from 
smaller scales due to the absence of  $k^3$ in the denominator of the integrant in Eq~(\ref{eq:bl_nonlin}), 
compared with $b_{{\rm L} \ell}(r)$. 
Since the perturbations in photon fluid becomes smaller at large $k$ for isocurvature perturbations,
the $b^{\rm (iso)}_{{\rm NL} \ell}$ is not as large as $2b^{\rm (adi)}_{{\rm NL} \ell}$.
This changes Eq.~(\ref{eq:f_TNL}) obtained by using the approximated transfer function in
the Sachs-Wolfe regime Eq.~(\ref{eq:SWtrans}).
We will discuss this issue below.


\begin{figure}[t]
 \begin{center}
   \includegraphics[width=1.0\linewidth]{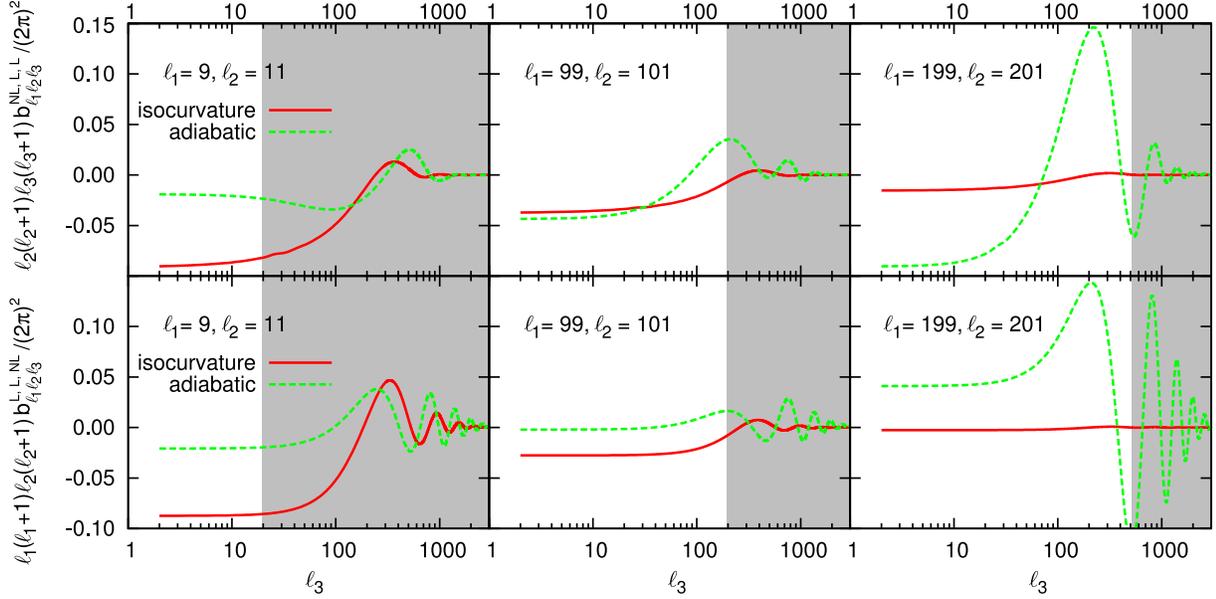}
   \caption{   
   The reduced bispectra 
   $\ell_2(\ell_2+1)\ell_3(\ell_3+1)b^{\rm NL,\ L,\ L}_{\ell_1 \ell_2 \ell_3}/(2\pi)^2$ (top)
   and 
   $\ell_1(\ell_1+1)\ell_2(\ell_2+1)b^{\rm L,\ L,\ NL}_{\ell_1 \ell_2 \ell_3}/(2\pi)^2$ (bottom).
   To avoid complexity, we have fixed $(\ell_1, \ell_2)=(9,11)$ (left), $(99,101)$ (middle), $(199,201)$
   (right) and varied $\ell_3$.
   The solid red line and dashed green line correspond to the cases with
   isocurvature and adiabatic initial conditions, separately. The unobservable multipoles are shown as shaded region.
   We have set $f_S = 1$.
   }
   \label{fig:nl}
 \end{center}
\end{figure}


Among the three terms in Eq.~(\ref{tem7}), we denote the two of them by
\bea
b^{\rm NL,\, L, \, L}_{\ell_1 \ell_2 \ell_3} &\equiv &
\int dr r^2 b_{{\rm NL}\ell_1}(r)b_{{\rm L}\ell_2}(r)b_{{\rm L}\ell_3}(r),\non\\
b^{\rm L,\, L, \, NL}_{\ell_1 \ell_2 \ell_3} &\equiv& 
\int dr r^2 b_{{\rm L}\ell_1}(r)b_{{\rm L}\ell_2}(r)b_{{\rm NL}\ell_3}(r).
\eea
In Fig.~\ref{fig:nl}, we show the bispectrum $b^{\rm NL,\, L, \, L}_{\ell_1 \ell_2 \ell_3}$
and $b^{\rm L,\, L, \, NL}_{\ell_1 \ell_2 \ell_3}$ for the isocurvature
and adiabatic perturbations.
One can see that the amplitude of the
bispectrum of the isocurvature perturbations is enhanced at large angular scales 
($\ell_1, \ell_2, \ell_3\lesssim 10$), compared with the adiabatic perturbations.
Our numerical calculations give the ratio at large scales as
\begin{equation}
 \frac{b^{\rm (iso)}_{\ell_1\ell_2 \ell_3}}{b^{\rm (adi)}_{\ell_1\ell_2 \ell_3}}\;\simeq \;4.
 \label{eq:bl3ratio}
\end{equation}
If the transfer function $g^{\rm (iso)}_{T\ell}(k)$ in Eq.~(\ref{eq:SWtrans}) were valid at smaller scales beyond the Sachs-Wolfe plateau,
the right hand side of Eq~(\ref{eq:bl3ratio}) would be $8$, as expected from Eqs.~(\ref{tem8}) and (\ref{eq:ratiot}).
However as have mentioned above, 
$b_{{\rm NL} \ell}(r)$ receives contributions from smalle scales ($k\gtrsim 1/\sqrt{3}\tau_*$) in a destructive way,
and this roughly halves the the amplitude of the bispectrum from isocurvature perturbations compared to
that from adiabatic perturbations.
At acoustic region ($\ell_1, \ell_2, \ell_3\simeq 200$), 
the amplitude of $b^{\rm (iso)}_{\ell_1 \ell_2 \ell_3}$ becomes much suppressed since CMB temperature anisotropies
are small at acoustic region for isocurvature initial perturbations.


\begin{figure}[t]
 \begin{center}
   \includegraphics[width=0.8\linewidth]{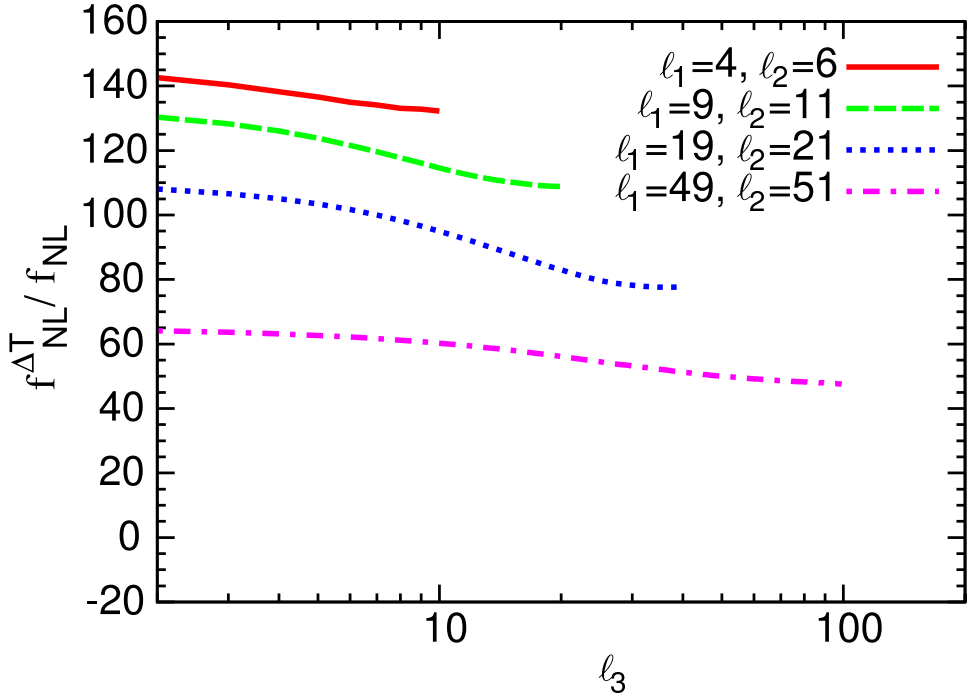}
   \caption{$f^{\Delta T}_{\rm NL}/f_{\rm NL}$ is plotted as a function of $\ell_3$ with various sets
   of $(\ell_1, \ell_2$). Only observable multipoles ($|\ell_1-\ell_2| \leq \ell_3 \leq \ell_1+ \ell_2$) are shown.}
   \label{fig:feff}
 \end{center}
\end{figure}


In Fig.~\ref{fig:feff} we plot $f^{\Delta T}_{\rm NL}/f_{\rm NL}$ as a function of $\ell_3$ for various sets of $(\ell_1,\ell_2)$.
We can see that our estimate of $f^{\Delta T}_{\rm NL}$ at large angular scales in Eq.~(\ref{eq:f_TNL}) is corrected
and approximately given by
\begin{equation}
	f_{\rm NL}^{\Delta T}\approx 100 f_{\rm NL}.
\end{equation}
However  the amplitude of the bispectrum at large angular scales
is still enhanced for non-Gaussian isocurvature perturbations compared with adiabatic perturbations.
In the adiabatic case, the largest signals of the primordial non-Gaussianity come from
the acoustic regions. On the other hand, in the isocurvature case, the signals 
are concentrated in the large angular scales. This remarkable difference in 
$\ell$-dependence of $b^{\rm (iso)}_{\ell_1 \ell_2 \ell_3}$ and $b^{\rm (adi)}_{\ell_1 \ell_2 \ell_3}$ will help 
us distinguish the  isocurvature non-Gaussianity from the adiabatic one.

So far, we have assumed SCDM universe. Here we give some comments on how the above results
will change in the standard $\Lambda$CDM universe. Assuming the flat universe, 
the amount of CDM is smaller ($\Omega_{c}\simeq 0.21$ \cite{Komatsu:2008hk}) in the $\Lambda$CDM universe.
For the adiabatic initial perturbations, this does not cause much difference on the CMB anisotropies at
large angular scales, except for the small late-time Integrated Sachs-Wolfe effect from 
nonzero $\Omega_\Lambda$.
On the other hand, for the isocurvature initial perturbations, 
the anisotropies at large angular scales will be  suppressed.
This is because the universe is not completely matter-dominated at recombination and 
therefore the curvature perturbations on large scales 
are not entirely generated from isocurvature perturbations in CDM 
(Eq.~(\ref{zeta-S}) is not a very good approximation in the $\Lambda$CDM model).
At large angular scales, the CMB anisotropies with the isocurvature initial perturbations are, however, 
still larger than for adiabatic ones and our discussion above is basically valid even in the $\Lambda$CDM 
universe. When constraining the isocurvature non-Gaussianity by using the future CMB data, 
we should take more realistic $\Lambda$CDM model,  but 
there will be no fundamental difference.
 
\section{Application to the axion} \label{axion}

In this section we apply our formulation to the axion as a concrete example.
The axion, $a$, is a pseudo Nambu-Goldstone boson associated with the spontaneous breaking of
the  PQ symmetry. Let us denote the breaking scale by $F_a$, whose magnitude is constrained from
various experiments, astrophysical and cosmological considerations.
The most strict lower bound on $F_a$ comes from the observation that
the duration of the neutrino burst in SN1987A lasted for $\sim $10 seconds.
In order to prevent too fast cooling by the axion emission, 
$F_a\gtrsim 10^{10}~$GeV is required~\cite{Raffelt:1990yz}.
On the other hand, the upper bound is provided by the cosmological argument.
The axion obtains a tiny mass after the QCD phase transition due to the anomaly effect which
explicitly breaks the PQ symmetry. The axion begins to oscillate coherently after that.
Since the lifetime of the axion is very long, it survives until now and contributes to DM of the universe.
The abundance is estimated as \cite{Turner:1985si}
\begin{equation}
	\Omega_a h^2 \;\simeq\; \left \{
	\begin{array}{ll}
	\ds{0.2\left ( \frac{F_a \theta^{1.7}}{10^{12}~{\rm GeV}} \right )^{1.18}}
	&~~~{\rm for}~~~F_a \theta > H_{\rm inf}/2\pi, \\
	 &\\
	\ds{0.2\left ( \frac{F_a}{10^{12}~{\rm GeV}} \right )^{-0.82}
	\left ( \frac{H_{\rm inf}/2\pi}{10^{12}~{\rm GeV}} \right )^{2}}
	&~~~{\rm for}~~~F_a \theta < H_{\rm inf}/2\pi,
	\end{array} \right. 
\end{equation}
where $\theta$ denotes the initial misalignment angle of the axion. Imposing 
$\Omega_a h^2 < \Omega_c h^2 \sim 0.11$~\cite{Komatsu:2008hk},
we obtain an upper bound on the PQ scale as $F_a \lesssim \theta^{-1.7}10^{12}$~GeV.
Thus the ratio of the axion abundance to the total dark matter abundance
$r \equiv \Omega_a/\Omega_{\rm CDM}$ is calculated as
\begin{equation}
	r \;\simeq\; 1.8 \times 10^{-2}\left ( \frac{F_a}{10^{12}~{\rm GeV}} \right )^{-0.82}
	\left ( \frac{a_*}{10^{11}~{\rm GeV}} \right )^2
	\left ( \frac{0.11}{\Omega_{\rm CDM}h^2} \right ),
\end{equation}
where we have defined
\begin{equation}
	a_* \;\equiv\; {\rm max} \left \{ F_a \theta,~~\frac{H_{\rm inf}}{2\pi} \right \}.  \label{a*}
\end{equation}

If the PQ symmetry is already broken before or during inflation and if it is never restored after inflation,
the axion has  unsuppressed quantum fluctuations during inflation because it remains 
practically massless during inflation.
Since the axion  contributes to some fraction of DM,
such an axionic isocurvature fluctuation is converted to the CDM isocurvature fluctuation.
Thus the axion is a plausible candidate for generating the non-Gaussianity 
from the isocurvature perturbation.

Now let us estimate the magnitude of the axionic isocurvature fluctuation and 
the resultant non-Gaussianity.
We assume that the inflaton itself does not generate non-Gaussianity, 
and that only the axion has an isocurvature fluctuation.
The axion acquires a quantum fluctuation during inflation given by
\bea
	\langle \delta a_{\vec k_1} \delta a_{\vec k_2} \rangle&=&(2\pi)^3
	\delta(\vec k_1+\vec k_2)P_{\delta a}(k_1),\non\\
	\Delta_{\delta a}^2(k) &\equiv&
	\frac{k^3}{2\pi^2}P_{\delta a}(k)=\left(\frac{H_{\rm inf}}{2\pi}\right)^2.
\eea
The observationally relevant quantity is the CDM isocurvature perturbation, rather than
the axionic isocurvature perturbation itself.
Using Eq.~(\ref{iso2}), the CDM isocurvature perturbation $S$ is given by
\begin{equation}
	S \simeq  \frac{\delta \rho_a}{ \rho_{\rm m 0}+ \rho_{a 0} },
\end{equation}
where $\rho_{\rm m0}$ denotes the dark matter abundance other than the axion.
In this equation, $\delta \rho_a$ is evaluated
on the uniform density slicing. Since $S$ is conserved quantity as long as the 
scales of interest are sufficiently large, we can evaluate it when the axion starts to oscillate.
When the axion starts to oscillate,
the universe is dominated by the radiation, and therefore
we can safely neglect the density fluctuation of the radiation, $\delta \rho_r/\rho_r$. 

As will become clear later, if one imposes the current constraint on the isocurvature
perturbation, large non-Gaussianity is generated only for $r \ll 1$.
Using $r \ll 1$,  we can approximate $S$ as
\begin{eqnarray}
S\; \simeq \; r \bigg[ \frac{2a_i \delta a}{a_*^2}+{\left( \frac{\delta a}{a_*} \right)}^2 \bigg], \label{SCDM}
\end{eqnarray}
where $a_i=F_a\theta$ denotes the classical deviation from 
the potential minimum.
The first term in (\ref{a*}) dominates over the second one when the classical deviation 
from the potential minimum
overcomes the amplitude of the quantum fluctuation.
In the opposite case, i.e., the isocurvature fluctuation is dominated by 
the second term in (\ref{SCDM}),
the whole dynamics of the axion is controlled by the quantum fluctuation
generated during inflation.
This is similar to the ``ungaussiton'' scenario~\cite{Suyama:2008nt}:
the axion is predominantly produced by the quantum fluctuations, 
giving the non-Gaussianity to the density fluctuations, 
while its contribution to the total curvature perturbation is negligibly small.

For later convenience, the power spectrum of the isocurvature perturbation $P_S(k)$ is defined through Eq.~(\ref{Psij}),
\begin{equation}
	\langle S_{\vec k_1} S_{\vec k_2} \rangle\;=\;(2\pi)^3
	\delta(\vec k_1+\vec k_2)P_{S}(k_1).
\end{equation}
Also note that the isocurvature perturbation is uncorrelated with the primordial curvature perturbation
in the case of the axion,
\begin{equation}
	\langle S_{\vec k_1} \zeta^{\rm inf}_{\vec k_2} \rangle \;=\;0.
\end{equation}

It is straitforward to calculate the non-linearity parameter $f_{\rm NL}$
defined by Eq.~(\ref{defnl}).
First note that from Eqs.~(\ref{fSij-1}) and (\ref{fSij-2}) $f_S$ is calculated as
\begin{equation}
	f_S \;\simeq\; \left \{
	\begin{array}{ll}
	{\ds \frac{1}{2r}}
	&~~~{\rm for}~~~F_a \theta > H_{\rm inf}/2\pi \\
	&\\
	{\ds \frac{1}{2r}\left( \frac{{a_*}}{{\Delta_{\delta a}}}\right)^2 \frac{1}{\ln (kL)}
	=\frac{1}{\Delta_\zeta}\left ( \frac{P_\zeta(k)}{P_S(k)} \right )^{1/2}[\ln (kL)]^{-1/2}}
	&~~~{\rm for}~~~F_a \theta < H_{\rm inf}/2\pi
	\end{array} \right. ,
\end{equation}
where $\Delta_\zeta^{2}\simeq 2.5\times 10^{-9}$ is the WMAP normalization of the 
curvature perturbation.
Then Eq.~(\ref{fNL}) tells us that the non-linearity parameter is given by
\begin{equation}
	f_{\rm NL}=\frac{5}{324r} \left ( \frac{P_S(k)}{P_\zeta(k)} \right )^{2},  \label{fNLbound1}
\end{equation}
if the classical deviation overcomes the quantum fluctuation 
$(F_a \theta > H_{\rm inf}/2\pi)$.
Since the WMAP five-year results give a contraint $P_S/P_\zeta \lesssim 0.1$,
a small value of $r \lesssim 10^{-4}$ is necessary for large non-Gaussianity, $f_{\rm NL} \gtrsim O(1)$.
But there is an upper bound on the level of non-Gaussianity coming from CDM isocurvature perturbation. 
In order to maximize the value of $f_{\rm NL}$, the isocurvature perturbation must also be large.
However, since $S$ is limited as $S < r$, $f_{\rm NL}$ is maximized at $2r \sim 0.3 \Delta_\zeta$
in order to saturate the isocurvature bound and this gives 
a strict upper bound as $f_{\rm NL} \lesssim 20$.
This is explicitly shown in the case $\delta a \gg F_a\theta$, where the quantum fluctuation dominates
the axion dynamics, giving $S\simeq r$.
In this case the non-linearlity parameter is estimated as
\begin{equation}
\begin{split}
	f_{\rm NL}&=\frac{5}{162}|\Delta_\zeta|^{-1}\left ( \frac{P_S(k)}{P_\zeta(k)} \right )^{3/2}
	\left[ \ln (kL)\right ]^{-1/2}\\
	&\simeq 20 \left ( \frac{P_S(k)/P_\zeta(k)}{0.1} \right )^{3/2}
	\left [\ln (kL)\right ]^{-1/2}.    \label{fNLbound2}
\end{split}
\end{equation}
Note that in this regime the parameter dependence of $f_{\rm NL}$ is all compressed in 
the information of the magnitude of the isocurvature perturbation.
Thus non-Gaussianity parameter $f_{\rm NL}$ is solely bounded by the isocurvature constraint
in this regime.
In other words, $f_{\rm NL}$ is maximized when the isocurvature contribution saturates the
allowed maximum value and the bound $f_{\rm NL}\lesssim 20$ does not depend on
other model parameters.
It may also be useful to give a full expression for $f_{\rm NL}$, 
\begin{equation}
	\frac{6}{5}f_{\rm NL}=\frac{1}{27N_\phi^4} \left ( \frac{2r}{a_*^2} \right )^3 \left[ 
		 a_i^2 + \Delta_{\delta a}^2\ln (kL)
	\right],
\end{equation}
whose limiting behavior approaches to the expressions given above.

Above results can be understood in a simple way. 
The following rough estimations may be useful because they give a correct parameter dependence.
When $\delta a \ll F_a \theta$ the isocurvature perturbation is dominantly given by
the linear Gaussian part $S\sim S^{({\rm g})}$, 
and hence the three point function is estimated as 
$\langle SSS \rangle \sim f_S P_S^2 \sim f_{\rm NL}P_\zeta^2$. 
Thus we obtain
\begin{equation}
	f_{\rm NL}\sim \frac{1}{r}\left ( \frac{P_S}{P_\zeta} \right )^2.
\end{equation}
On the other hand when $\delta a \gg F_a \theta$, 
The non-Gaussian part dominates the isocurvature perturbation $S\sim f_S S^{({\rm g})2}$,
giving a three point function as $\langle SSS \rangle \sim P_S^{3/2} \sim f_{\rm NL}P_\zeta^2$. 
As a result we obtain
\begin{equation}
	f_{\rm NL}\sim \Delta_\zeta ^{-1} \left ( \frac{P_S}{P_\zeta} \right )^{3/2}.
\end{equation}

\begin{figure}[t]
 \begin{center}
   \includegraphics[width=0.8\linewidth]{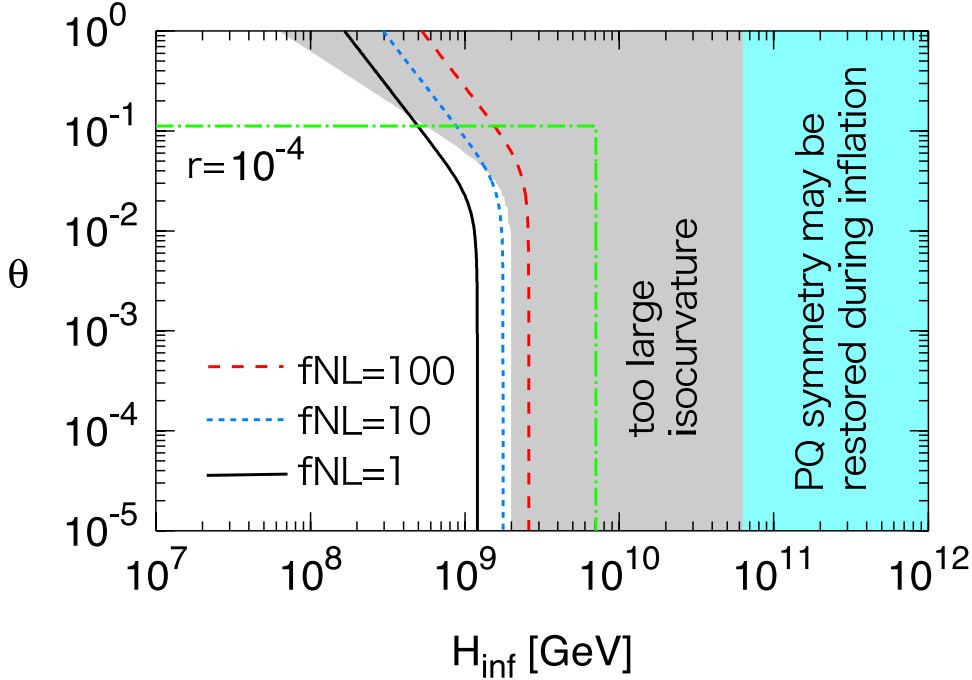} 
   \caption{
  	Contours of $f_{\rm NL}=1,10$ and $100$ for $F_a = 10^{10}~$GeV.
	Gray shaded region is excluded from isocurvature constraint.
	In the blue region the PQ symmetry may be restored during inflation, and so,
	neither isocurvature fluctuation nor non-Gaussianity will arise.
	Also we show $r=10^{-4}$ by the green dash-dotted line.
   }
   \label{fig:NG-Fa10}
 \end{center}
\end{figure}

\begin{figure}[htbp]
 \begin{center}
   \includegraphics[width=0.8\linewidth]{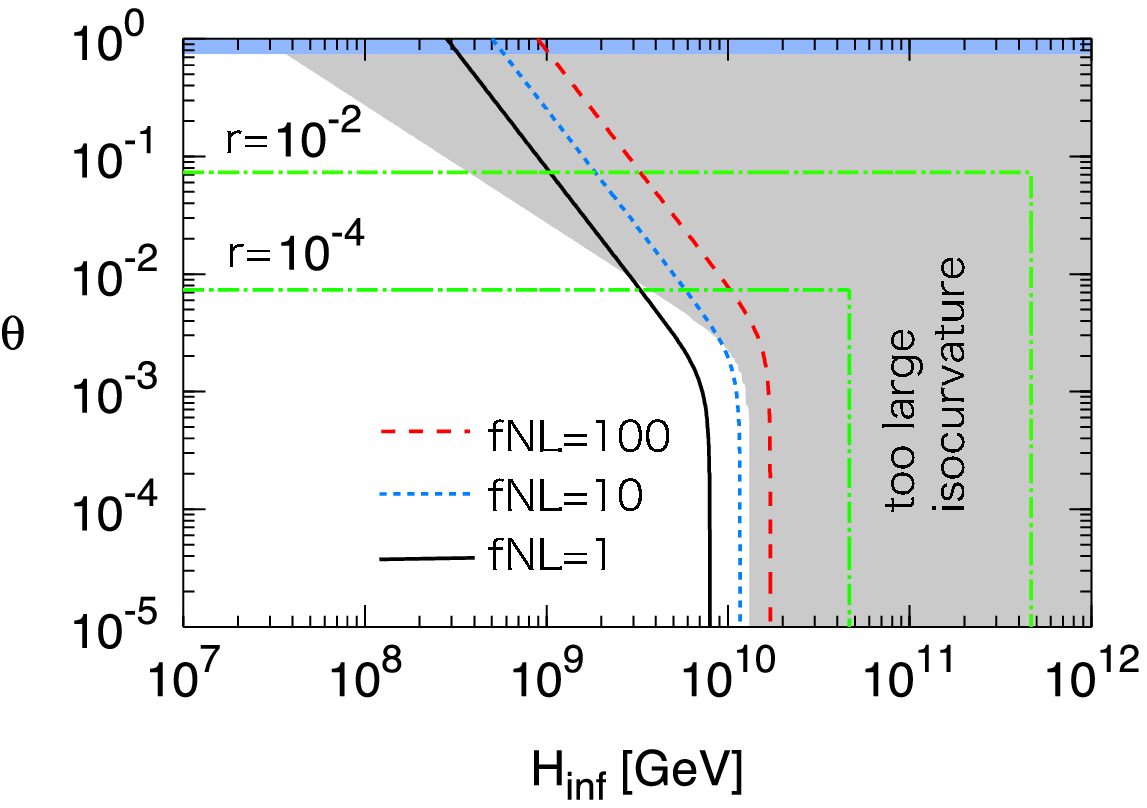} 
   \caption{
   Same as Fig.~\ref{fig:NG-Fa10} with $F_a = 10^{12}~$GeV.
   The upper shaded region is excluded from the axion overproduction.
   }
   \label{fig:NG-Fa12}
 \end{center}
\end{figure}

\begin{figure}[htbp]
 \begin{center}
   \includegraphics[width=0.8\linewidth]{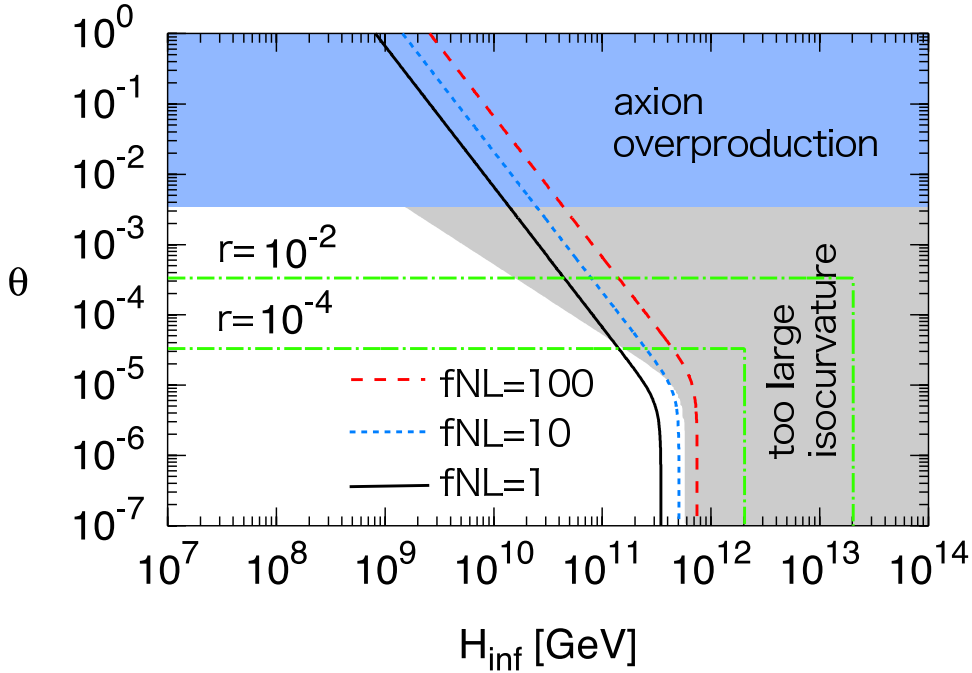} 
   \caption{
   Same as Fig.~\ref{fig:NG-Fa10} with $F_a = 10^{16}~$GeV.
   The upper shaded region is excluded from the axion overproduction.
   }
   \label{fig:NG-Fa16}
 \end{center}
\end{figure}

In Figs.~\ref{fig:NG-Fa10}, \ref{fig:NG-Fa12} and  \ref{fig:NG-Fa16},  the non-linearlity parameter $f_{\rm NL}$ is shown
on $H_{\rm inf}$-$\theta$ plane for $F_a = 10^{10}$~GeV, $10^{12}$~GeV, and  $10^{16}$~GeV.
It is seen that an observable amount of non-Gaussianity (say, $f_{\rm NL}\gtrsim O(10)$)
is generated near the isocurvature constraint.
However, notice that the relevant quantity from CMB observations is $f_{\rm NL}^{\Delta T}$,
not $f_{\rm NL}$, and the relation between them is given in Sec.~\ref{sec:temp}.
In particular it has been shown that $f_{\rm NL}^{\Delta T}$ can be $100$ times larger than $f_{\rm NL}$,
depending on the observed scale.
Thus it may be possible that isocurvature fluctuation is probed only through its 
non-Gaussian imprints on CMB.


\section{Conclusions and Discussion} \label{conclusion}
In this paper we have investigated a possibility that large non-Gaussianity is generated
by isocurvature fluctuations. One interesting feature of this scenario is that 
the bispectrum and the power spectrum of the CMB temperature fluctuations exhibit
characteristic scale dependence.
In particular, the effective non-linearity parameter $f_{\rm NL}^{\Delta T}$ is significantly enhanced
at large scales,  compared to the adiabatic case.
Furthermore, our results indicate that large non-Gaussianity may be accompanied 
with an observable fraction of the isocurvature perturbation.
If future observations confirm both large non-Gaussianity and 
a certain amount of isocurvature fluctuation component, our scenario will become very attractive. 
As a concrete example, we have shown that the axion can naturally induce such
isocurvature perturbation in the CDM sector, leading to large non-Gaussianity.
If the axion is indeed responsible for the large non-Gaussianity 
hinted by the current observations, the inflationary scale should be in the range of
$O(10^9 - 10^{11})$\,GeV. 
This opens up an interesting possibility that the axion can be probed through 
its non-Gaussianity contribution to the CMB temperature fluctuation, even if 
the energy density of the axion today is negligible compared to the total 
dark matter abundance~\footnote{
There may be an anthropic reason that forces the axion abundance to
take such value~\cite{NT}.
}.

Although we have restricted ourselves to the CDM isocurvature perturbation in this paper, 
the baryonic isocurvature perturbation can also generate large non-Gaussianity
in a similar fashion \cite{Kawasaki:2008jy}. 
There is indeed a scenario using the flat direction with the baryon number
to generate baryonic isocurvature perturbations~\cite{EM,KT,Kasuya:2008xp}.

We have assumed that the CDM isocurvature perturbation comes from the fluctuations in the CDM sector.
However, this may not be the case, and it can arise from the radiation. That is to say,
the CDM isocurvature perturbation is given by  $S=3(\zeta_{\rm CDM}-\zeta_r)=-3\zeta_r$ in some cases,
including the curvaton/ungaussiton scenario.  In this case the sign of the non-linearity parameter $f_{S}$ 
becomes negative, which is expected to leave distinct features on the bispectrum of the temperature fluctuation.
One may be able to use the features to probe into the origin of the isocurvature perturbation
\cite{Kawasaki:2008pa}.

In general, the isocurvature perturbation mainly affects the large scale temperature anisotropy,
while its effect on the density perturbation is weaker than the adiabatic one.
Thus if the non-Gaussianity is truly sourced by the isocurvature perturbation,
it can be seen only in the CMB observations, and we should have null detection of non-Gaussianity
from the analyses using the matter power spectra. Hopefully, 
future cosmological observations will enable us to establish or refute the existence of large
non-Gaussianity. If it is there, we may be able to distinguish the origin, adiabatic or isocurvature.
Undoubtedly, once detected, it will provide us with  useful information on the early universe and
the high energy physics.

\section*{Acknowledgment}

K.~Nakayama and T.~Sekiguchi would like to thank the Japan Society for the Promotion of
Science for financial support.
This work is supported by Grant-in-Aid for Scientific research from the Ministry of Education,
Science, Sports, and Culture (MEXT), Japan, No.14102004 (M.K.)
and also by World Premier International
Research Center Initiative (WPI Initiative), MEXT, Japan.



{}

\end{document}